\documentclass{mn2e}
\usepackage[dvips]{color}
\usepackage{epsf}

\title[V838 Mon photometric study]
{The post-outburst photometric behaviour of V838 Mon}
\author[L. A. Crause et al.]
{Lisa A. Crause,$^{1}$\thanks{E-mail: lcrause@artemisia.ast.uct.ac.za
(LAC); wal@ph.adfa.edu.au (WAL); dmk@saao.ac.za (DMK); fvw@saao.ac.za (FVW); 
fm@saao.ac.za (FM); afjones@ts.co.nz (AFJ)} 
Warrick A. Lawson,$^{2\star}$ David Kilkenny,$^{3\star}$ Francois van 
Wyk,$^{3\star}$ \and Fred Marang,$^{3\star}$ and Albert F. Jones$^{4\star}$\\
$^{1}$Department of Astronomy, University of Cape Town, Private Bag,
Rondebosch 7700, South Africa\\
$^{2}$School of Physics, University of New South Wales, Australian
Defence Force Academy, Canberra ACT 2600, Australia\\
$^{3}$South African Astronomical Observatory, P.O. Box 9, Observatory
7935, South Africa\\
$^{4}$Carter Observatory, P.O. Box 2909, Wellington, New Zealand\\}

\date{Accepted .......... Received .......... in original form ..........}

\pagerange{\pageref{firstpage}--\pageref{lastpage}}
\pubyear{2002}

\begin{document}

\maketitle

\label{firstpage}

\begin{abstract}

The unusual eruptive variable discovered in Monoceros in 2002 January
underwent dramatic photometric and spectroscopic changes in the months prior
to its 2002 June--August conjunction with the Sun.  Optical and infrared
(IR) photometry obtained at the South African Astronomical Observatory
(SAAO) between 2002 January and June (JD $2452280-440$) is presented here in
an analysis of the star's post-outburst behaviour.  The light curve
indicated 3 eruptions took place in 2002 January, February and March. SAAO
\'echelle spectra obtained in the week prior to the March maximum indicated
the ejection of a new shell of material. {\it JHKL\,} photometry obtained
during 2002 April showed the development of an IR excess due to the
formation of a dust shell.  The shell appears to be largely responsible for
the rapid fade in the optical flux during 2002 April--May ($\Delta V > 6$
mag within 3 weeks).  Blueing of the optical colours during the decline is
likely due either to the revealing of an emission line region surrounding
V838 Mon, or the unveiling of the progenitor or a spatially-close early-type
star.

\end{abstract}


\section{Introduction}

Eruptive stellar events provide rare opportunities to test and refine
theoretical models, particularly those involving short-lived evolutionary
phases.  These brief episodes afford valuable insight into the physical
changes that stars undergo when transforming from one state to another.
V4334 Sgr (better known as Sakurai's Object; Duerbeck et al. 2000) is the
best recent example of such a variable.  The star was `caught in the act' as
it underwent a late helium shell flash in 1996 and has shown the evolution
of post-AGB stars passing through this phase to be faster than theory
predicted (Iben et al. 1983).  The wealth of information gathered from this
event demonstrated the value of intensive multi-wavelength monitoring of
rapidly-evolving variable stars for improving our understanding of the late
stages of stellar evolution.  With similar motivations, the peculiar
nova-like object in Monoceros (now known as V838 Mon) has been observed
regularly since it was reported in outburst in 2002 January (Brown 2002)
until its conjunction with the Sun in 2002 June.

A possible progenitor for V838 Mon was identified as a $V \approx 15.6$ star
by Munari et al. (2002) from POSS-I plates and the 2MASS IR sky survey.  The
spectral energy distribution derived from these data suggest the progenitor
was possibly an F-type star with moderate interstellar reddening
[$E$($B-V$) $\approx 0.5$ mag].  In the months following the outburst,
optical and IR light curves were obtained for V838 Mon across a broad ($\sim
100$ d) maximum that was complex in nature, suggesting multiple outburst
events.  Assuming the F-star was the progenitor, the $V$ amplitude of the
initial outburst was $\sim 6$ mag and comparison between 1997 2MASS data and
2002 January SAAO IR data shows {\it JHK\,} magnitudes brightened by $\sim
7$ mag.  The amplitude from quiescence to maximum was almost 10 magnitudes
in both the optical and IR.  Spectroscopically the star appeared to
transform from a mid-K type giant into a late-M supergiant between 2002
January and May (Munari et al. 2002). The Balmer series, Na I D lines and
various low ionisation species exhibited P-Cygni profiles during the various
maxima, while at other times H$\alpha$ showed either pure absorption or pure
emission.  CCD imaging revealed a light-echo beginning to expand around the
star about two weeks after the February maximum (Henden, Munari \& Schwartz
2002) and subsequent observations have shown the reflection nebula continuing
to grow in size and evolve in form.

Early studies of V838 Mon likened it to classical novae, final helium shell
flash stars such as Sakurai's Object, a mysterious eruptive variable in M31
and the `unusual luminous red variable' V4332 Sgr (Munari et al. 2002). 
While it has certain properties in common with each of these classes or
individual objects, departures from their characteristics are significant
and the nature of V838 Mon remains unclear.  On-going monitoring of
this object is a priority given the dramatic changes observed during the six 
months following the initial outburst.

Optical and IR photometry of V838 Mon was obtained with several SAAO
telescopes between 2002 January and June.  Multi-filter CCD images of the
expanding light-echo were also obtained with the SAAO 1.0-m telescope during
this time, as well as post-conjunction.  These data will be presented
elsewhere (Crause et al. in preparation)\footnote{({\it VRI\,}) colour
images were produced from these data; see the Astronomy Picture of the Day
website at {\tt http://antwrp.gsfc.nasa.gov/apod/ap021003.html}.} in a
discussion of the star's distance.  In this paper we present the SAAO
photometry, merged with visual estimates obtained by one of us (AFJ) and
optical/IR photometry reported previously by Munari et al. (2002).

\section{Observations and Data Reduction}

Cousins {\it UBVRI\,} optical photometry was obtained with the SAAO 0.5-m
and 1.0-m telescopes.  Aperture photoelectric photometry was performed for 
all 0.5-m data, while most of the 1.0-m observations were made with the 
1k $\times$ 1k STE4 CCD.

The aperture photometry was reduced with standard SAAO software, with the
measurements of V838 Mon calibrated against Cousins E-region standards.  The
CCD frames were bias-subtracted, overscan-trimmed and flat-fielded using the
{\tt ccdproc} package within {\tt IRAF}.  The clean images were then reduced
with the {\tt phot} package in which one is able to specify the extraction
radius to exclude the nebula from the aperture.  Although the nebula was not
present in CCD frames obtained in 2002 January, these data were reduced in
the same way as later images for consistency.  A set of comparison stars was
measured in each frame and calibrated against Landolt equatorial standards
observed on the 3 best nights of the 2002 April-May observing run.  A mean
magnitude was calculated for each of the comparison stars and these were
then used to calibrate the differential magnitudes determined for V838 Mon. 
Finally the colour equation for the STE4 CCD was applied to produce the
magnitudes and colours presented in Table 1, however the extreme post-maximum 
colours [($V-R$) $\approx 3$ and ($V-I$) $\approx 6$] required large 
extrapolations for the standard transformations.

\begin{table}
\begin{center}
\caption{SAAO Cousins {\it UBVRI\,} photometry of V838 Mon obtained
from 2002 January to June.}
\begin{tabular}{cccccc}
JD--2452000 & $V$ & ($B-V$) & ($U-B$) & ($V-R$) & ($V-I$) \\
\hline
289.388  &  9.72  &  2.01  &  --    &  --    &  2.22  \\
290.436  &  9.94  &  1.71  &  1.82  &  0.93  &  1.85  \\
293.419  & 10.40  &  1.82  &  2.02  &  0.99  &  1.92  \\
294.401  & 10.12  &  1.82  &  1.97  &  0.98  &  1.92  \\
295.417  & 10.12  &  1.79  &  1.96  &  0.97  &  1.91  \\
311.300  &  6.87  &  1.08  &  0.22  &  0.70  &  1.45  \\
311.427  &  6.84  &  1.07  &  0.22  &  0.70  &  1.45  \\
313.285  &  7.00  &  1.13  &  0.45  &  0.70  &  1.46  \\
314.296  &  7.21  &  1.16  &  0.62  &  0.70  &  1.47  \\
315.350  &  7.41  &  1.18  &  0.68  &  0.70  &  1.47  \\
316.284  &  7.58  &  1.22  &  0.74  &  0.72  &  1.49  \\
316.289  &  7.58  &  1.22  &  0.74  &  0.71  &  1.49  \\
318.422  &  7.80  &  1.33  &  --    &  0.79  &  1.60  \\
319.398  &  7.84  &  1.40  &  --    &  0.81  &  1.64  \\
320.458  &  7.87  &  1.47  &  --    &  0.84  &  1.68  \\
321.368  &  7.90  &  1.56  &  --    &  0.84  &  1.70  \\
326.408  &  8.00  &  1.81  &  1.32  &  0.98  &  1.94  \\
327.321  &  8.03  &  1.86  &  1.35  &  1.01  &  1.98  \\
331.304  &  8.16  &  2.04  &  1.56  &  1.08  &  2.09  \\
335.279  &  7.96  &  1.96  &  1.59  &  1.04  &  2.04  \\
336.296  &  7.83  &  1.89  &  1.53  &  1.02  &  1.99  \\
337.376  &  6.26  &  1.85  &  1.50  &  0.99  &  1.92  \\
389.317  & 11.75  &  2.18  &  1.68  &  2.49  &  6.12  \\
390.236  & 12.19  &  2.14  &  1.52  &  2.61  &  6.00  \\
392.226  & 13.04  &  2.02  &  1.16  &  2.90  &  6.49  \\
393.313  & 13.36  &  --    &  --    &  3.05  &  6.70  \\
395.269  & 13.90  &  1.76  &  0.88  &  3.01  &  6.86  \\
396.241  & 14.28  &  1.66  &  --    &  3.16  &  7.12  \\
397.241  & 14.49  &  1.57  &  --    &  3.18  &  7.16  \\
398.231  & 14.65  &  1.46  &  --    &  3.16  &  7.23  \\
399.227  & 14.83  &  1.38  &  --    &  3.18  &  --    \\
428.224  & 16.17  &  --    &  --    &  2.22  &  6.70  \\
\hline
\end{tabular}
\end {center}
\end{table}

{\it JHKL\,} photometry was obtained with the IRP Mark II photometer
on the SAAO 0.75-m telescope.  These data were reduced with standard SAAO
software, placing the magnitudes on the Carter (1995) system.  Most of the 
observations made during 2002 January and February were presented by Munari 
et al. (2002).  Subsequent observations, as well as an additional set of 
measurements made in January, are listed in Table 2.  The typical uncertainty 
is 0.02 mag for the {\it JHK\,} data and 0.05 mag for the {\it L}-band 
observations.

Visual observations of V838 Mon were obtained from soon after the initial
report of the outburst during 2002 January until the star's conjunction with
the Sun.  The visual estimates are in agreement at the 0.2 mag level with
contemporaneous photoelectric and CCD measurements at times when the colour
of the variable was not extreme or rapidly varying.  Once the star began to
fade and redden during April--May, both the uncertainty of the estimates and
the discrepancy with the photoelectric photometry increased; see Section 3.5.

\begin{table}
\begin{center}
\caption{SAAO {\it JHKL\,} photometry of V838 Mon obtained with
the 0.75-m telescope between 2002 January and June.  These data
are additional to SAAO IR photometry published by Munari et al.
(2002).}
\begin{tabular}{cccccc}
  JD--2452000      & {\it J} & {\it H} & {\it K} & {\it L} \\
\hline
  287.502  &  6.79  &  6.15  &  5.91  &  5.55 \\
  347.328  &  4.02  &  3.36  &  3.03  &  2.70 \\
  348.306  &  4.03  &  3.36  &  3.02  &  2.66 \\
  395.226  &  4.95  &  3.99  &  3.49  &  2.63 \\
  395.233  &  4.95  &  3.99  &  3.48  &  2.63 \\
  396.242  &  4.99  &  4.02  &  3.51  &  2.65 \\
  425.181  &  6.00  &  4.71  &  3.87  &  2.92 \\
  426.176  &  6.04  &  4.75  &  3.89  &  2.92 \\
  429.179  &  6.14  &  4.79  &  3.92  &  2.92 \\
\hline
\end{tabular}
\end{center}
\end{table}

\section{Photometric Analysis}

Optical and IR light and colour curves (Figs 1 and 2), as well as
colour-magnitude and colour-colour diagrams (Figs $3-5$) illustrate the
complex photometric behaviour of V838 Mon from 2002 January to June (JD
$2452280-440$).  The optical data consist of observations from SAAO, as well 
as those obtained with the 1.0-m telescope at the United States Naval 
Observatory (USNO), those from a 0.25-m telescope in Tsukuba, Japan 
(Munari et al. 2002), and AFJ's visual estimates.  The $B$-band Tsukuba 
observations are shown with a $-0.25$ mag offset applied so that these 
data agree with the SAAO and USNO $B$-band photometry.  All IR data are 
from the SAAO 0.75-m telescope. 

The Julian dates of key events in the light curve are given in abbreviated
form (JD$-2452000$) throughout the remainder of the text.  The principal
stages in the light curve are described below and are labelled (i) through
(v) in each figure: (i) is centred near JD 290 (mid-January) when the
brightness peaked for the first time; (ii) covers the interval JD
$300-330$, including the February maximum that occurred on JD 311; (iii)
covers JD $330-360$ including the March peak on JD 345; (iv) spans JD
$360-400$ to include the April fade and (v) is from JD 400 (early-May) until
V838 Mon's conjunc tion with the Sun in June when the star was at its faintest.  

\begin{figure}
\begin{center}
\epsfxsize=8.4cm
\epsffile{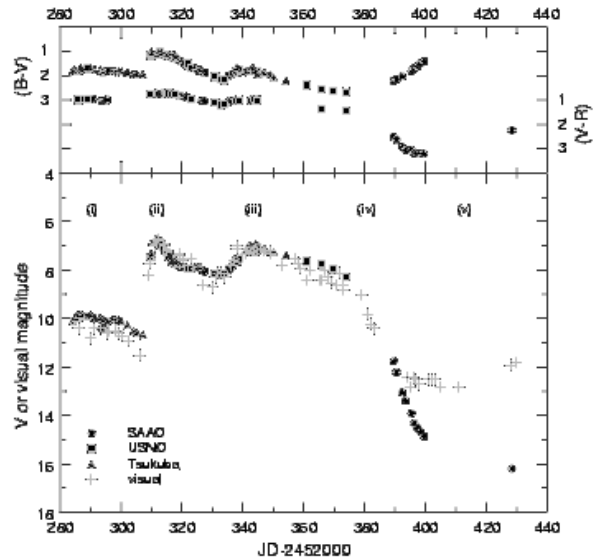}
\caption{($B-V$) and ($V-R$) colour curves (top and bottom respectively in
upper panel) and the {\it V\,} and visual light curve from the initial 
outburst in 2002 January (JD 280) through to the deep decline in May and 
June (JD 440).  Important stages in the light curve are labelled (i)--(v); 
see Section 3.}
\end{center}
\end{figure}

\begin{figure}
\begin{center}
\epsfxsize=8.4cm
\epsffile{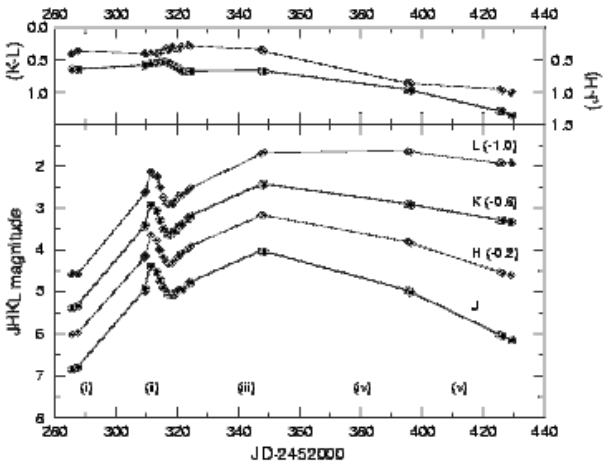}
\caption{($K-L$) and ($J-H$) infrared colour curves (top and bottom 
respectively in upper panel) and {\it JHKL\,} light curves for V838 Mon 
obtained between 2002 January and June.  All data are from the SAAO 0.75-m 
telescope with most of the stage-(i) and (ii) points also published by 
Munari et al. (2002).  {\it HKL\,} curves are offset for clarity by the 
amounts (in magnitudes) shown.}
\end{center}
\end{figure}

\begin{figure}
\begin{center}
\epsfxsize=8.8cm
\epsffile{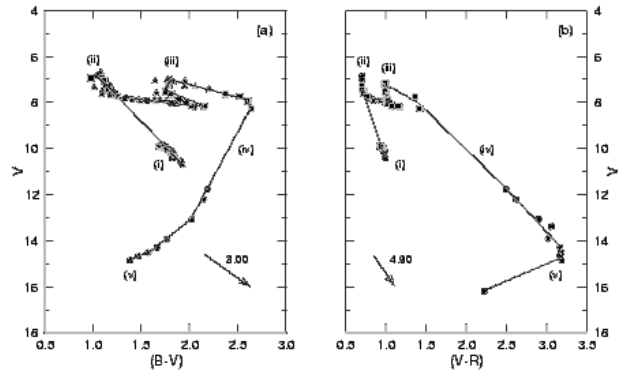}
\caption{Optical colour-magnitude diagrams showing the evolution of V838 
Mon between 2002 January and June.  For the symbol key, see Fig. 1. The 
reddening vectors are calculated from equations given by Bessell \& Brett 
(1988).}
\end{center}
\end{figure}

\subsection{(i) Initial Peak}

The progenitor of V838 Mon is believed to have had a {\it V\,} magnitude of
$\sim 16$ (Munari et al. 2002).  Applying the reddening of $E(B-V) = 0.7$
given by Kimeswenger et al. (2002) and using a distance of 2.3--2.5 kpc 
determined from the expansion of the light echo (Bond et al. 2002, Crause et 
al. in preparation) rather than the 790 pc derived by Munari et al., we find 
M$_{V} \sim 1.7$.  This brighter progenitor absolute magnitude is inconsistent 
with the `under-luminous F main sequence star' proposed by Munari et al.  The 
nova-like 2002 January eruption that led to the star's discovery (Brown 2002) 
reached $V \approx 10$ (M$_{V} = -3.9$) by JD 285.

It soon became apparent that the star was not a typical classical nova;
spectra showed unusually weak Balmer lines (Wagner, Halpern \& Jackson
2002), ejection velocities were of the order of 350 km\,s$^{-1}$ and the
post-outburst decay was uncharacteristically slow.  Such low velocities are
usually associated with slow novae (Allen 1980) and are expected in post
asymptotic-branch giant branch stars undergoing a late thermal pulse (Iben
et al. 1983), leading to early speculation (Della Valle \& Iijima 2002;
Henden et al. 2002; Rauch, Kerber \& van Wyk 2002) that the star may be
similar to `born again' stars such as Sakurai's Object and V605 Aql.

\subsection{(ii) Maximum Outburst}

Instead of continuing to decrease in brightness after the January maximum,
V838 Mon suddenly began a second outburst on 2002 February 3 (JD 309). 
Brightening by $\approx 0.1$ mag\,hr$^{-1}$ for 2 days (AAVSO and VSNET
observations), the star peaked at $V = 6.7$ on February 5, corresponding to
M$_{V} \sim -7.2$.  The now hotter star ($\sim 6500$ K according to
Kimeswenger et al. 2002) produced the blue maximum seen in the optical and
IR colour curves; however we note a lag of 4 days between the optical maximum 
and the peak in the ($J-H$) colour.  The latter may be related to the 
appearance of the second overtone $^{12}$CO ($\Delta v = 3$) absorptions 
in the {\it H}-band (Banerjee \& Ashok 2002).  Over the 3 weeks following the 
maximum the star faded to $V \approx 8$ while the ($B-V$) colour reddened by 
$\approx 0.9$ magnitudes.

A spectrum obtained just before the February outburst was classified by
Zwitter \& Munari (2002) as that of a `heavily reddened cool K-type star'. 
At maximum, spectra were dominated by narrow P-Cygni profiles and the Balmer
lines, particularly H$\alpha$, had increased in strength (Iijima \& 
Della Valle 2002).  This eruption also resulted in a light-echo that was 
first observed on 2002 February 17 (JD 323; Henden et al. 2002).

\begin{figure*}
\begin{center}
\epsfxsize=15.0cm
\epsffile{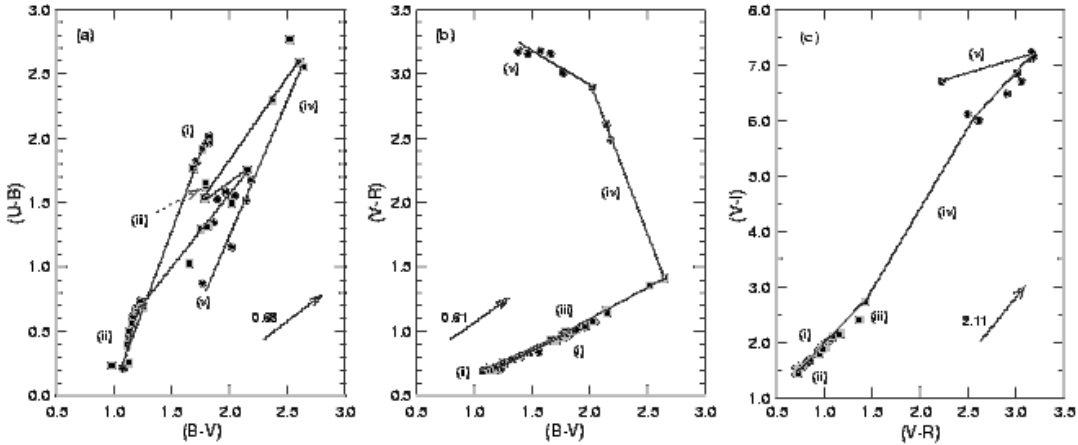}
\caption{Optical colour-colour diagrams illustrating the blue outbursts, 
followed by a reddening phase and finally the `blue decline' as the star 
rapidly faded.}
\end{center}
\end{figure*}

\subsection{(iii) March Peak}

In early March, V838 Mon brightened for a third time, peaking at $V \approx
7$ on 2002 March 11 (JD 345) and producing another blue maximum in the
($B-V$) colour curve.  Comparison between curve-(iii) and curve-(ii) in Fig. 
6 shows the star fading in the optical but remaining bright at IR wavelengths, 
the lag noted in Section 3.2.  Throughout the remainder of March and 
early-April the star again faded, more gently than after the previous two 
outbursts, and the optical colours reddened again (see Fig. 1).

Sample high-resolution spectra obtained with the SAAO fibre-fed \'echelle
spectrograph on the 1.9-m telescope (resolution $R = 32,000$; Crause et al.
in preparation) during the week prior to this maximum are shown in Fig. 7. 
As the spectra are yet to be corrected for the star's radial velocity,
features are described relative to the rest wavelength of H$\alpha$.  The
original P-Cygni profile that appeared during the February outburst steadily
weakened and is seen to distort as a second lower-velocity absorption
feature developed $\sim 60$ km\,s$^{-1}$ blueward of the rest wavelength of
H$\alpha$.  This profile appears to indicate the ejection of a new shell of
material, despite the relatively low amplitude ($\Delta V$ $\sim 1.5$) of
this eruption.  Part of the February absorption component can still be seen
$\sim 160$ km\,s$^{-1}$ blueward of this new feature.  {\it JHK} spectra 
obtained just prior to the March maximum were reported by Geballe et al.
(2002) to show a combination of CO absorption and emission, indicating both 
photospheric and circumstellar components.

\subsection{(iv) IR Excess Development and Reddening}

From mid-April (JD $\sim 380$) V838 Mon began a sharp decrease in
brightness, fading $\Delta V > 6$ mag within 3 weeks.  The ($V-R$) and
($V-I$) colours reddened significantly (by $\sim 1.5$ and 4.0 mag,
respectively) while the ($B-V$) colour showed a strong blueward trend
[$\Delta$($B-V$) $= -1.0$ mag; also see Section 4.2 for a discussion of the
origin of the blue excess].  The latter is best seen in the ($B-V$)/$V$
colour-mag diagram (Fig. 3a) where the movement in the ($B-V$) colour was
nearly perpendicular to the reddening vector, as well as in in Fig. 4(a)
which shows the evolution in the ($B-V$)/($U-B$) colour-colour plane. 
Curve-(iv) in Fig. 6 also clearly shows the development of a blue excess. 
The ($B-V$)/($V-R$) plane (Fig. 4b) shows the combination of a blueing
($B-V$) and a redward-moving ($V-R$) while in the ($V-R$)/$V$ and
($V-R$)/($V-I$) diagrams (Figs 3b and 4c, respectively) the trend in the
colours during the decline is similar to pure reddening.  The change in the
spectral type of the star from mid- to late-M (Henden et al. 2002) during 
this time must also be in part responsible for the colour variations.  Using 
the reddening relations of Bessell \& Brett (1988) we conclude that reddening
accounts for most of the decline phase of the light curve.  If we assume the
entire change in colour [$\Delta$($V-I$) $\approx 4$ mag; see Fig. 4] is due
to the effects of reddening then we obtain an upper limit for
$A_{V} = 9.5$ mag, which is similar to the observed $V$-band amplitude. 
This value for $A_{V}$ overestimates the effects of reddening since the star
continued to evolve towards later spectral types during the decline. 
However, as the star was spectral type $\approx$ M5 near the mid-point of
the decline (JD 385; Rauch et al. 2002), the intrinsic colour of the star
could not have greatly exceeded ($V-I$) $\approx 4$ at this time
(Schmidt-Kaler 1982).

The reddening and dust formation scenario is supported by the IR light and
colour curves and colour-colour diagrams.  The IR colours shifted redward as
the contribution from the star at shorter wavelengths fell away (the $J$-
and $H$-band photometry will largely follow the stellar flux) while the $L$
magnitude (sensitive to warm circumstellar dust) remained essentially
constant.  In Fig. 5 we see that, following the loops in the colour-colour
diagrams due to lags in the IR colour changes during the three outbursts at
maximum, the star reddened rapidly [$\Delta$($K-L$) = 0.6 mag within 50
days] as it faded.  During the decline (stage iv and v), we find the ratio 
$\Delta$($V-I$)/$\Delta$($J-H$) $\approx 6$, which we compare to a ratio of
3.4 given by Bessell \& Brett (1988) for the case of reddening alone producing 
the decline.  As we discuss above, reddening appears to be the significant
(though not the only) contributor to the light and colour curve behaviour of
V838 Mon during the decline from outburst.  A lack of IR observations during
the fade from maximum prevents us from accurately determining the IR
behaviour at this time and so we simply connect the points from JD 347 (the
last SAAO IR data obtained at maximum) with those near JD 395 (when the star
was $V = 14$) in Figs 5a and 5b.  These lines have slopes of $\approx 0.6$
and 0.3, respectively, or only $\approx 0.25$ the slope of the reddening
vector in these colour-colour planes.  This colour behaviour strongly
indicates the development of an IR excess during the fade that most affects
the $K$- and $L$-band photometry.  After accounting for underlying reddening,
the additional IR excess due to the dust gives rise to a slope of $\approx
0$ in Fig. 5b.  

Banerjee \& Ashok (2002) report complex multicomponent structure in the CO 
bands during this time and confirm the persistence of both the absorption and 
emission components noted previously by Geballe et al. (2002).  They also 
identify a number of Ti I emission lines that first appeared in April 9 
{\it K}-band spectra and use these highly unusual features, produced in the 
circumstellar ejecta, to estimate the mass of the shell ($10^{-7}$ to 
$10^{-5}$ $M_{\odot}$).

\subsection{(v) Deep Decline}

Shortly before its June-August conjunction, V838 Mon had faded to $V \approx
16$.  With only {\it VRI\,} photometry available due to the faintness of the
star in the $U$ and $B$ bands, we could not follow the evolution of the blue
optical colours, but we note that by this time the ($V-R$) and ($V-I$)
colours had also developed a blueward trend [see stage (v) in Figs 1, 3b, 4c
and 6].

The star faded slightly in the IR during stage (v), more so at $J$ than at
$L$.  In Fig. 5(a) we see that the star appeared to undergo reddening during
this stage, while in Fig. 5(b) the slope was steeper than the reddening
vector, implying that the ($H-K$) excess was increasing relative to the
($K-L$) excess.  The strong water vapour absorption bands seen at 1.4 and
1.9 $\mu$m during May (Banerjee \& Ashok 2002) may be responsible for
these changes.  Banerjee \& Ashok also note that the Ti I emission lines
that appeared in the {\it K}-band in early-April peaked in strength during 
early-May and had decreased by mid-May.  While these lines would have 
influenced the IR colours, without {\it L}-band spectra we are unable to
assess the effect they had on the colour-colour diagrams in Fig. 5.  

The influence of the changing nature of the star on the visual photometry
can clearly be seen in stage (v).  Visual estimates were substantially
brighter than nearby CCD observations, by 4 mag when the star was at its
faintest.  Since the human eye retains sensitivity beyond the red cutoff of
{\it V}-band filters, the star's extreme ($V-I$) colour ($> 6$) resulted in
a substantial increase in the perceived visual magnitude above the observed
{\it V}-band measurements.  This effect was likely compounded by the
increasing contribution from the light-echo to the integrated flux of the
star and nebula, as well as the higher airmass of these observations.

\section{Discussion}

\subsection{The dust}

Kimeswenger et al. (2002) reported the presence of a weak 10 $\mu$m excess
that originated prior to the second maximum.  This excess increased over the
next several month and is consistent with our interpretation of new dust
having formed and subsequently cooled.  Kimeswenger et al. assume a
condensation temperature of 750 K and calculate that dust could form at a
distance of $1.4 \times 10^{9}$ km during February when the star was at its
hottest.  If we assume a velocity of 350 km\,s$^{-1}$, material ejected
during the February outburst would have reached this distance about 45 days
later, suggesting that dust formation could have begun in late March.  For a
short-lived dust-formation episode with the material being driven away from
the cooling star, we expect the dust temperature to decrease rapidly.  2002
September (JD $\sim 535$) data from Watson \& Costero (2002) show the mid-IR
excess peaking at about 8 $\mu$m, indicating a dust temperature of about 360
K.  To estimate the time it would take for the material to cool to this
extent, we make the simple assumption of constant shell luminosity and find
that the shell's radius would have to increase by a factor of $\sim 4$.  At
a velocity of 350 km\,s$^{-1}$ this would take about 200 days, leading to JD
555.  Since the star's evolution towards later spectral types would promote
cooling, we consider the 200 days to be an upper limit and so JD 555 is in
good agreement with the date of the mid-IR observations (JD 535).

\subsection{The blue excess}

The development of a blue excess when the star rapidly faded resembles the
`blue decline' behaviour often seen when R Coronae Borealis (RCB) variables
(Clayton 1996) enter an obscuration minimum.  In RCB stars this is thought
to be the result of a condensing dust cloud eclipsing the photosphere,
temporarily revealing a rich `chromospheric' emission line region (Cottrell,
Lawson \& Buchhorn 1990).  A similar mechanism may be responsible for the
colour behaviour seen in V838 Mon, although post-conjunction spectroscopy
suggests this excess is due to an early-type star (Wagner \& Starrfield 2002) 
which may the `progenitor' reported by Munari et al. (2002) or a 
spatially-close field star.

\subsection{Similar stars} 

Munari et al. (2002) and Kimeswenger et al. (2002) have discussed the
classes or objects that V838 Mon may or may not be related to.  Early
indications that the star may be a classical nova or a final helium shell
flash object now appear to have been ruled out by a number of factors.  The
current impression is that V838 Mon may belong to a new class of eruptive
variables, along with two other poorly-understood, rapidly-evolving red
objects known as M31 RV (Rich et al. 1989) and V4332 Sgr (Martini et al. 
1999).  

V4332 Sgr appears to be the closest analogue of V838 Mon in terms of its
general photometric behaviour and rapid redward spectral evolution.  This
object cooled from 4400 K to 2300 K in three months and the spectral type
changed from K3 III-I to M6 III-I in only a week.  While the light curve was
not multi-peaked, V4332 Sgr was only discovered to be in outburst when it
emerged from conjunction with the Sun and so it may too have undergone
earlier eruptions.  Martini et al. (1999) suggest that a nuclear event in a 
single star, in which a slow shock drove the photosphere outwards, could have 
powered the luminosity evolution and the emission spectrum of V4332 Sgr. 
Such a slow shock could also account for the principal difference between V838
Mon and V4332 Sgr, namely that post-outburst spectra of V4332 Sgr showed {\it 
inverse} P-Cygni profiles, perhaps indicating the infall of previously ejected 
material. 

Kimeswenger et al. (2002) exclude M31 RV from the proposed new class of
variables because of the extremely high luminosity (M$_{bol} = -10$) this
star achieved at maximum, compared to their M$_{V}$ $\approx -4.5$ for V838
Mon.  If, however, the larger distance estimates for V838 Mon [2.5 and 2.3
kpc respectively by Bond et al. (2002) and Crause et al. in preparation]
prove correct, the resulting M$_{V}$ $\approx -7$ would significantly reduce 
this difference between M31 RV and V838 Mon.  M31 RV was discovered in outburst
in 1988 and archival plates show that the red variable exhibited a similar 
maximum in 1968 (Sharov 1990).  This object should thus be monitored towards 
the end of the decade in the event that it erupts again as periodic outbursts 
would strongly support a binary scenario.  The rapid radial velocity variations 
(a  change of $\approx 400$ km\,s$^{-1}$ in just 8 hr) noted by Rich et al. 
(1989) should also be anticipated if the star erupts again.

\section{Summary}

We have presented optical and IR photometry for V838 Mon following the 2002
January outburst until its conjunction with the Sun in June.  Optical data
show multiple maxima followed by a reddening phase and then a rapid fade 
which was accompanied by blueing of the ($B-V$) and ($V-R$) colours.  The 
IR colours progressed redward during the decline and indicate a combination 
of reddening and the development of a ($K-L$) excess during this time. 
Spectra obtained shortly before the final maximum in March show the ejection 
of an additional shell of material.  The nature of V838 Mon remains unclear, 
although it appears to be closely related to the unusual V4332 Sgr which 
showed comparably dramatic evolution in 1994, and may be similar to M31 RV if 
larger distance estimates prove correct.

\begin{figure}
\begin{center}
\epsfxsize=8.8cm
\epsffile{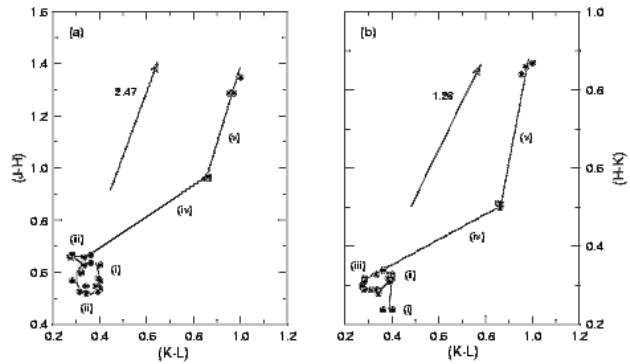}
\caption{Infrared colour-colour diagrams showing the development of a ($K-L$) 
excess and reddening of the star.}
\end{center}
\end{figure}

\begin{figure}
\begin{center}
\epsfxsize=8.4cm
\epsffile{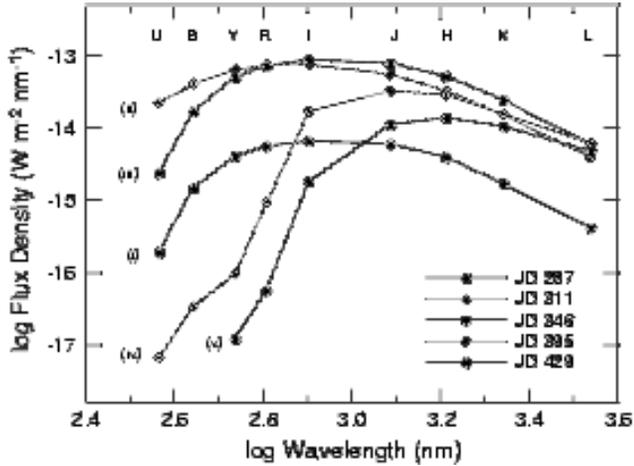}
\caption{Spectral energy distributions for V838 Mon during each of the five
stages discussed in Section 3.  The development of a blue excess as the
star faded can be seen in (iv) and to a lesser extent in (v) which lacks 
{\it U\,} and {\it B\,} data.  The much slower decrease in the IR, particularly 
towards longer wavelengths, is clearly visible in (ii)--(v).}
\end{center}
\end{figure}

\begin{figure}
\begin{center}
\epsfxsize=6.0cm
\epsffile{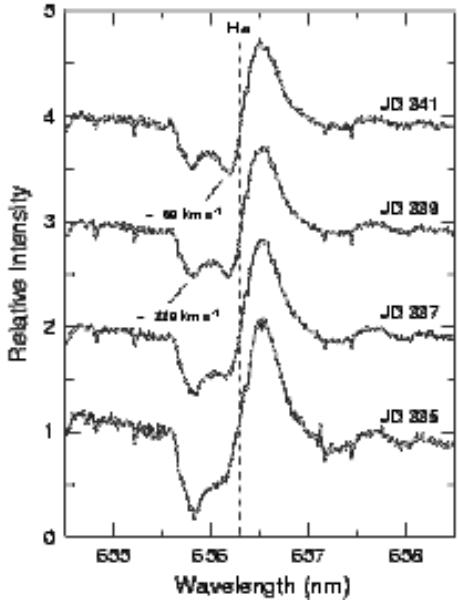}
\caption{2002 March SAAO \'echelle spectra showing the development of a second 
absorption feature ($\sim 60$ km\,s$^{-1}$ blueward of the H$\alpha$ rest 
wavelength) in the P-Cygni profile during the week prior to the maximum on JD 
345.  The remainder of the higher-velocity absorption feature from the February 
outburst can be seen further blueward, about $- 220$ km\,s$^{-1}$ from the 
H$\alpha$ rest wavelength.}
\end{center}
\end{figure}

\section*{Acknowledgements}

We thank Nye Evans for his constructive referee's report, Michael Feast
(University of Cape Town) for valuable discussions about these data, Chris
Koen (SAAO) for his useful comments on the manuscript, Wolfgang Zima
(University of Vienna) for obtaining the \'echelle data in March, Luis
Balona (SAAO) for assistance in reducing the spectra and John Menzies (SAAO)
for obtaining the JD 429 $VRI$ photometry.  LAC acknowledges support from
the National Research Foundation (NRF) through the grant-holder bursary
scheme (grant-holder Tony Fairall) and the University of Cape Town.  WAL
acknowledges research support from UNSW@ADFA Faculty Research Grants and
Special Research Grants.

\end{document}